\documentclass{iopjournal}

\usepackage{amsmath}
\usepackage{booktabs}
\usepackage{multirow}
\usepackage{makecell}

\hypersetup{allcolors=black}

\begin{document}

\articletype{Paper}

\title{Gaze-controlled segmentation improves object recognition accuracy in simulated phosphene vision}

\author{Eleftherios Papadopoulos$^{1,*}$, Umut G\"{u}\c{c}l\"{u}$^{1}$ and Ya\u{g}mur G\"{u}\c{c}l\"{u}t\"{u}rk$^{1}$}

\affil{$^{1}$Donders Institute for Brain, Cognition and Behaviour, Radboud University, Nijmegen, The Netherlands}

\affil{$^{*}$Author to whom any correspondence should be addressed.}

\email{lefteris.papadopoulos@donders.ru.nl}


\begin{abstract}
  \textit{Objective.} Visual neuroprosthetics offer a promising solution, providing a simplified visual sense through devices comprising cameras, computers, and implanted electrodes. A central challenge is deciding what to convey with the few phosphenes available. We asked whether letting the user's gaze select the object that is conveyed in full, rather than extracting edges across the whole scene, improves object recognition in simulated phosphene vision.

  \textit{Approach.} We developed Gaze-Controlled Segmentation (GCS), in which the point of gaze prompts the Segment Anything Model and the selected object is rendered as a filled region over the scene's edges at reduced intensity. In a simulated phosphene-vision environment, eight participants performed an object search-and-localization task under GCS, edge detection, and normal-vision conditions. Both phosphene conditions were rendered gaze-contingently, and the data were analysed with Bayesian mixed-effects models with participants and images as crossed random effects.

  \textit{Main results.} Relative to edge detection, GCS produced a modest but credible improvement in identification accuracy on target-present trials ($+9.4$ percentage points, 95\% highest-density interval $[+4.3, +14.2]$). Gaze reached the target about as often under both methods, consistent with an advantage at the identification stage. GCS trials tended to be slower ($+2.7$~s, $[-0.8, +6.6]$), a difference that was not credible, and in exploratory analyses the advantage was largest for cylindrical objects and absent for round ones. Stated preferences leaned toward GCS on most usability criteria, although with eight participants these tendencies are individually consistent with chance.

  \textit{Significance.} The comparison is between the integrated GCS representation and edge rendering, and does not isolate the contribution of the segmentation itself. These findings illustrate the potential of gaze-driven, interactive scene representation for neuroprosthetic vision, while underscoring that the gain is modest, task-dependent, and established so far only in simulated phosphene vision with sighted observers, using phosphenes larger than current implants would produce.
\end{abstract}

\section{Introduction}

For individuals around the world, visual impairments create difficulties in carrying out everyday activities. Studies have documented their prevalence, with tens of millions affected by blindness and hundreds of millions experiencing moderate to severe visual impairments globally \cite{Bourne2021}. In addition, the impact of blindness or limited vision has been studied extensively, highlighting the reduction in quality of life \cite{STELMACK2001,Kamelska2015} and the array of associated social challenges, like social stigmatization \cite{Bulk2020}, isolation \cite{Rokach2021}, poverty and unemployment \cite{Gold2005}. An emerging and promising method to potentially aid visually impaired people against these issues are visual neuroprosthetics \cite{Fernandez2018,Beauchamp2020}. Recent scientific milestones include encouraging results from animal testing \cite{Chen2020} and human trials \cite{Holz2026}\cite{Allen2025}\cite{Fernndez2021} that hold promise for the safety \cite{niketeghad2020phosphene} and efficacy \cite{muqit2024prosthetic} of these devices. They evoke localized light percepts known as phosphenes, whose position and intensity can be modulated to convey a coherent representation of the visual scene \cite{Schmidt1996}.

Despite the exciting prospect of these devices, several important limitations need to be addressed. One of these limitations is that due to the low amount of electrodes, the phosphene representations that can be elicited are of very low resolution. The necessary amount of electrodes that need to be implanted in order to restore functional vision is estimated to be around  625 \cite{Cha1992}. Comparing this number with the 44 to 378 electrodes implanted in recent human trials \cite{Fernndez2021}\cite{Holz2026}\cite{Allen2025}, it becomes apparent that the current phosphene vision resolution achievable is far from ideal. It is also challenging to determine the appropriate stimulation parameters to ensure that what users perceive aligns with researchers' intentions while maintaining their safety---an issue that ongoing research is actively addressing \cite{Kkolu2025}.     Moreover, while the colour of phosphenes cannot be sufficiently controlled, their shape also varies across individuals, further limiting representational capacity \cite{EricksonDavis2021,Beyeler2019}. Given these limitations, it is important to emphasize that visual neuroprosthetics cannot restore healthy vision but only provide a simpler form of visual sensation due to the limited information that can be represented. This is the reason why current research focuses on developing efficient algorithms that can transform the camera's natural images to a phosphene representation that is best understood by the user \cite{deRuytervanSteveninckend}.

The earliest and still most common family of approaches extracts low-level features, such as edges \cite{Guo2018,Vergnieux2017}. This approach is favored by many researchers because it requires minimal computational resources, making it optimal for portable computers that users can carry with them. In \cite{Thorn2020} an extensive set of experiments was conducted, including object identification and road crossing in simulated environments, while allowing participants to explore the scene with head and eye movements. Similarly, in \cite{deRuytervanSteveninck2022}, participants had to move in a real environment where head movements and body movements affected the phosphene representation of the scene. A second family conveys semantic rather than geometric structure. Semantic segmentation has gained traction in recent years due to the development of robust deep learning models \cite{Han2021,SanchezGarcia2020}.  With this approach, researchers are able to convey not just the outline of an object but also its semantic identity, adding a significant amount of information to the phosphene representation. In some cases these two techniques have been combined by extracting edges based on the semantic categories of objects. In \cite{Rasla2022} semantic edges were combined with depth cues to test obstacle avoidance and object recognition. While combining the two methods was important for obstacle avoidance, neither the combination nor the individual methods performed well for object recognition, and participants did not prefer semantic edges. In \cite{Kasowski2025} semantic edges were used by displaying all task-relevant edges at once  or sequentially over time in a wayfinding task. This paper highlights the importance of semantic identity as a tool to achieve less cluttering compared to traditional edge detection approaches. In most of these approaches the algorithm decides, in advance and in the same way for every user, which parts of the scene the phosphenes convey. Where implant users have any control, it is a switch between whole-image processing modes, such as edge detection or contrast inversion \cite{EricksonDavis2021} or near and far distance filters \cite{Kartha2020}, although it has been argued that the most useful representation depends on the task and that users should be able to change it on demand \cite{Beyeler2022}. Beyond neuroprosthetics, a broad low-vision human--computer interaction literature augments residual vision for closely related tasks: visual cues that steer attention during a visual search \cite{Zhao2016}, visual and audio wayfinding guidance delivered through smartglasses \cite{Zhao2020}, and camera-based close-range navigation support that detects and guides users toward objects \cite{Hao2023}. These systems share the central design goal of our work---reducing a scene to the information the user actually needs---and motivate using gaze as the mechanism that selects what is rendered.

Despite the variety of approaches used to create a simplified visual sense, several problems remain unaddressed. First, since not every research group has access to actual implant users, most scientists use phosphene simulators to replicate phosphene vision for sighted individuals.  However, their methods do not always achieve the same results in actual device users \cite{Beyeler2019,EricksonDavis2021}, emphasizing the need for more realistic simulators \cite{Dagnelie2008}. Additionally, it has been shown that eye movements play a significant role in understanding a scene with phosphene vision \cite{Sabbah2014}, an element often omitted in the evaluation of algorithms. Because phosphenes move with the eyes while a head-mounted camera does not, the region of the camera image sent to the implant has to follow the measured gaze position \cite{Paraskevoudi2019,Caspi2018}, and in simulation such gaze-contingent processing improved mobility, scene recognition and visual search, mostly by removing this spatial conflict \cite{deRuytervanSteveninck2024}. Gaze-contingent rendering is therefore a requirement for a realistic simulation, and it determines where the scene is sampled rather than what the phosphenes convey. Lastly, we identify a gap in research related to utilizing the gaze signal, which such a system tracks in any case, to let users actively select what the presented phosphene representation conveys in a way that promotes the users' capacity to explore their environment. Addressing this gap is timely given recent evidence that implant users' own priorities frequently diverge from the assumptions of the researchers who design for them \cite{Nadolskis2024}, which argues for representations that are to an extent determined by the user, rather than fixed in advance by the algorithm alone.

In the current study we implemented a Gaze-Controlled Segmentation (GCS) algorithm. It keeps the gaze-contingent rendering of \cite{deRuytervanSteveninck2024} and uses the gaze position a second time, as the prompt to the Segment Anything Model \cite{Kirillov}, a promptable segmentation foundation model whose generality could greatly improve the overall system. Rather than assigning semantic labels, our algorithm lets users highlight and fill in the shape of an object by directing their gaze toward it, conveying object form through interactive selection. To test its efficiency, we designed an object search experiment, utilizing a state-of-the-art biologically plausible simulator \cite{vanderGrinten2024} and evaluated the time needed to complete the task, object identification accuracy and gaze behaviour. The simulator we used models stimulation of the primary visual cortex, and is to our knowledge the most detailed phosphene model that is publicly available. We used it for its realistic layout of phosphenes and their gaze-contingent behavior, while the size and brightness of the phosphenes followed stimulation settings chosen for visibility rather than for clinical realism (Section 2.6). The proposed method is not specific to one implant site and could in principle be applied to retinal and thalamic implants as well. Participants' preferences between the two methods were collected with a short survey at the end of the experiment. Following \cite{Bollen2019}, who debriefed participants on their experience and preference between two phosphene rendering methods, we formalised this into a forced choice between the two methods on six criteria (overall preference, clarity of object appearance, visual comfort, eye fatigue, mental effort and visual appeal). We hypothesize that GCS, in which the gazed object is filled and the remaining edges are carried at reduced intensity, improves accuracy over edge detection but may increase the time required to complete the task.  We also expected participants to display significantly different exploration strategies between the two methods. Lastly, the method is to be considered as a whole and we do not distinguish between the effects of the different components of our method.

\section{Experiment}

\subsection{Participants}

We recruited nine participants (3 male, 6 female) with a mean age of 24.9 years (SD 1.4). To be included in the study, participants had to have normal or corrected to normal vision and no ocular malformations. None of them wore glasses or contact lenses during the experiment. None of the participants had prior experience with phosphene vision simulations. All participants gave written informed consent and the study was approved by our institution's ethical committee.

All nine met the inclusion criteria and one was excluded, leaving eight. This participant's results were negative towards our hypothesis, but the exclusion was made because the eye-tracker calibration was visibly off during the session, something the participant themselves reported. The recorded gaze carries a large, directional offset through the first half of their gaze-controlled session and, separately, through half of their edge-detection session, so neither condition was delivered to them as intended.

No prospective power or precision calculation was performed. The sample size was set by the practical constraints on data collection within the project, and is comparable to sample sizes reported in the simulated-prosthetic-vision literature.

\subsection{Experimental Setup}

The participants were seated in front of a screen with a resolution set at 1920x1080 pixels. At a viewing distance of 50 cm from a 24-inch monitor, the display subtended approximately $56^\circ$ of visual angle horizontally and $33^\circ$ of visual angle vertically. We used a chin rest to stabilize their head during the experiment which was positioned appropriately so that the participant's eyes would align with the top quarter of the monitor. We also used an Eyelink 1000 eye tracker to track the participants' gaze. The device was positioned properly so that the distance from the eye-tracker to the chinrest is approximately 50 cm and to ensure it is centered horizontally to the monitor. Then the height of the device was adjusted so the eye-tracker was positioned as high as possible without blocking the view of the display.

The eye-tracker was set to binocular tracking at 500 Hz. Our aim was to have the participants control a point on the screen with their gaze. This gaze point was used as the prompt to the segmentation model; it was not displayed to participants during the experiment. It was visible during the normal-vision training session in order to get familiar with how the eye tracking works. In order for the point to lie closer to where participants experience their center of vision, we averaged the positions of the two gaze coordinates taken from the two eyes. To optimize the tracking fidelity, we reduced the power of the illuminator to 75\% which reduces the amount of reflections on the cornea. During the calibration phase of the eye-tracker, we set the background to be black and the calibration point to be white.

\subsection{Experimental process}

In our experiment, participants were asked to locate requested objects when presented with images. These images were taken from the ARID10 and ARID20 subsets of the OCID dataset \cite{DBLP:conf/icra/SuchiPFV19} which is prominently used for robotic vision research. A typical sample from ARID consists of  a cluttered scene consisting of everyday objects on the floor or on a table. All images are taken from the same point of view. An example of such an image is shown in Figure \ref{fig:base}. The goal was to compare their accuracy and behavior between three conditions, the GCS condition (our method), the Edge Detection condition and the normal-vision condition.

Before the start of the experiment, participants were introduced to both methods with the same slide presentation: one scene outside the experimental set, shown as a photograph, under GCS and under Edge Detection, with an animation of how each rendering follows gaze. Additionally, they were shown images of all possible target objects and were asked to familiarize themselves with them in order to set their expectations to a specific shape for each target. This step was incorporated based on feedback from a pilot study participant who noted that their differing expectations regarding the object's shape increased the difficulty of the experiment. Lastly, the participants did a practice run of six trials under the normal-vision condition, on images not used in the experiment, in order to familiarise themselves with using their gaze to control the display. Neither GCS nor Edge Detection was practised in its own rendering.

\subsection{Experimental Design}

Our design was inspired by previously conducted object search experiments \cite{Parikh2013,Nejad2025}. The experiment consisted of three sessions, each using a different image processing method and representing the three conditions. During each session the participants were presented with 76 prompts consisting of a black screen and the name of a requested object and an equal amount of processed images. The same 76 image-target pairs were presented in all three sessions. A single random presentation order was generated for each condition and used for every participant. The selection of targets was made under the constraint that there were not two objects that would be indistinguishable without further features like colour. For example, if an image contained two types of fruit similar in shape, like one orange and one lemon, neither of those could be a target. Both this constraint and the familiarisation with the target images described above make the task easier than open-ended object recognition and limit generalisation accordingly. No shape-confusable alternative was present in any scene, and participants held a specific shape expectation for each target before the session began. The accuracy reported below should therefore be read as an upper bound relative to a setting in which the target is unknown in advance and objects of similar shape may compete with it. Furthermore, the images were grouped into three sets according to the number of objects existing in the scene which will be referred as cluttering. These three sets were low cluttering with 1 to 3 objects, intermediate cluttering with 5 to 8 objects and high cluttering with 9 to 12 objects. Low cluttering contained 24 samples while intermediate and high cluttering contained 26 samples. In the GCS and Edge Detection conditions, there was a brief break after 38 images so the participants could rest. There was no break in the normal-vision condition. We presented the first 4 recruited participants with the GCS condition before the Edge Detection condition and the remaining participants with the opposite ordering. All participants were presented with the normal-vision condition last. The normal-vision condition served as a reference for the scale of the difference between phosphene and normal vision rather than as a condition entering the method comparison; it was therefore run last and was not counterbalanced with respect to the two phosphene conditions. Calibration was done before the start of each condition. All three sessions were completed by each participant on the same day with between-session breaks of at most 5 minutes.

Participants were expected to search for the object in the image using their gaze. After finding the object, they were asked to move the mouse pointer onto it and press the left mouse button to report its position. The mouse pointer was drawn as a crosshair, that is, two thin perpendicular lines meeting at the pointer position, so that participants could see exactly where a click would land; it was drawn on top of the displayed image and was the only pointer shown. In some trials the requested object was not included in the image. In these cases participants were asked to press the right mouse button, anywhere on the image. These two button presses were the only responses available, so each trial ended with a single mouse response that reported both whether the requested object was present and, when it was, where it was. We refer to these trials as false trials, while trials in which the requested object was included in the image are referred to as true trials. Each session contained 54 target-present and 22 target-absent trials.

After the end of the experiment, participants had to fill a short questionnaire regarding their experience under the GCS and Edge Detection conditions. As for the data collection, we stored the coordinates of the points the participants clicked at, denoting where they thought the requested object was, the time from the presentation of the processed images to their decision and the trajectory of their gaze.

\subsection{Image processing}

We applied three different image processing strategies, two of which were the main conditions we wanted to contrast, the GCS condition and the Edge Detection condition. The normal-vision condition included the raw images resized to 640x640. For the Edge Detection condition, we utilized the Canny Edge Detection algorithm from the scikit-image python library \cite{Van_der_Walt2014-kp}. We left the hysteresis thresholds at the scikit-image defaults of 25 and 50, and set the Gaussian smoothing parameter to $\sigma$ = 5 for the Canny edge detector. The smoothing parameter was the only one we chose. The first author determined the setting by visual inspection across the full image set, the criterion being that object outlines stay visible while floor and table texture drop out. It was fixed before data collection, never changed, and never adjusted in response to participant performance. A retrospective sweep over eight smoothing values and four threshold pairs, scored against the ground-truth object boundaries, puts this setting at the maximum of the grid and at the precision-recall crossover. The full grid is given in the Supplementary Material (Section S1). Then, we processed the image with the phosphene simulator around the current gaze point (Section 2.6), resized it to 640x640 and displayed it in the 640x640 area at the center of the screen.

For the GCS condition we utilized the SAM model. The SAM model is a promptable segmentation model. The user can select a point in the image, specified by its coordinates, and the model will segment the object encompassing that point. Our main idea was to use the point representing the center of the participant's gaze as the prompt from which the model extracts the mask of the object being looked at, and to highlight that mask. The participant therefore directs their gaze towards an area in the scene and the model determines which mask that point selects, filling that object and showing it with phosphenes.

Despite our efforts to implement the SAM processing in real-time, the limitations of the available hardware posed substantial challenges. We were unable to achieve a video frame rate higher than 14 frames per second.  To address this issue, we opted for pre-calculating the masks for each image and making them available for access during the experiment.

For the generation of the pre-calculated masks, we initially needed to preprocess the images to increase the quality of the output. Initially, every image was resized to 1024x1024 to fit the input resolution of the SAM model. The next step involved normalizing the histogram of the brightness for each image utilizing tools from the open-cv python library \cite{opencv_library}. To achieve this, we first converted the image to the YUV color space and equalized the histogram of the Y component (luma) which is associated with the illumination intensity of the image. Without this step, the Automatic Mask Generator frequently returned masks split along illumination boundaries --- a single object appearing as two or more masks divided by a shadow edge --- and masks that extended across an object's border into similarly lit background. Histogram equalisation of the luma channel substantially reduced both failure modes on visual inspection of the mask sets. This assessment was qualitative: no quantitative comparison of mask quality with and without preprocessing was performed, and the preprocessing was fixed before data collection, identically for all images. Finally, we convert the image back to the RGB color space and give it as input to SAM. Since processing time was not an issue anymore, we selected the largest model published by the developers which provided the highest quality segmentation results. Next, we used the Automatic Mask Generator functionality which calculates the masks of all objects in the scene. Our output consisted of a Nx1024x1024 binary matrix where N is the number of masks in a specific image identified by the model.

To simulate the mask extraction in real time, we initially loaded the appropriate mask matrix for each trial. We iterate over the N different masks and select those that have a true value at the gaze coordinate for each time point. This way we effectively highlight the shape of the object the participant was looking at. To provide information about objects outside the participant's direct gaze, we superimposed the Edge Detection image, scaled to approximately 30\% of the intensity of the segmentation mask.  As a last step, the image was processed and displayed as in the Edge Detection condition. The 640x640 pixel stimulus was presented centrally and subtended approximately 20.1 degrees of visual angle. The remainder of the display was black.

\subsection{Phosphene Simulator}

The last processing step of both Edge Detection and GCS conditions was to process the image using the phosphene simulator developed in \cite{vanderGrinten2024}. This implementation allows for biologically plausible control of the size and brightness of the phosphenes using the device's parameters like the electrical stimulation intensity and the stimulation frequency as well as controlling the position of the implanted electrodes. We followed the configuration files provided in \cite{githubGitHubNeuralcodinglabdynaphos}. Specifically, for the electrode positions we selected 600 electrode positions probabilistically according to the reverse dipole visuotopic model for the V1 area. The simulator's visual angle parameter was set to $4^\circ$. This parameter sets the extent of the simulated field and does not correspond to the visual angle of the phosphenes on the screen, which is given below. The pulse frequency was set at 300Hz following \cite{Fernndez2021}. The electric current for the electrode grid was set to $600\mu A$ for full-brightness regions and to 30\% of that for the reduced-brightness edges in GCS. This current is about nine times the average current at which implant users detected phosphenes \cite{Fernndez2021}, so these settings should be read as choices that set what was displayed rather than as a realistic stimulation protocol. Owing to an error in our implementation, the simulator was reset before each frame, so each frame showed the phosphenes at the onset of stimulation, without the build-up of brightness over successive frames that the simulator models. The remaining parameters were maintained as specified in the configuration files, with the exception that thresholding was disabled.

Phosphene size and brightness follow equations 4 to 10 of \cite{vanderGrinten2024} (Dynaphos version 0.1.3). The simulator's output ranges from 0 to 1 per pixel, with the summed phosphenes clipped at 1, which we scaled to represent gray levels from 0 to 255 without further normalization. This output depends on the mask given to the simulator but also on the position of each pixel relative to the phosphene's center. A value of 1 was shown as the monitor's full white. For example, the center of a single phosphene had a gray level of about 250 at full brightness and about 170 to 210 for the reduced-brightness edges in GCS, so the edges appeared only somewhat dimmer than the selected object. Single phosphenes were about 0.3$^\circ$ wide near the gaze point and about 0.8$^\circ$ wide 8$^\circ$ away from it (full width at half their peak), and phosphenes driven by the GCS edges were about half as wide. The monitor was not calibrated for luminance.

Both phosphene conditions were rendered gaze-contingently. On every update, the image was shifted so that the current gaze point was at its center, passed through the simulator, and shifted back, so the phosphenes moved on the screen with the eyes while the scene image stayed fixed. The simulator set each phosphene from the brightest pixel within its receptive field, using the receptive-field sampling of its configuration files \cite{githubGitHubNeuralcodinglabdynaphos}. With the used simulator parameters the phosphenes covered a circle approximately 20$^\circ$ in diameter centered on the gaze point, with about half of them within 3.5$^\circ$ of it. The display was updated at about 30 Hz with GCS and 40 Hz with Edge Detection, on a 60 Hz monitor. We did not time individual frames, but from these rates we estimate that a frame took 25 to 33 ms to draw, and at most about 38 ms in the slowest trials (26 frames per second), to which the eye tracker adds about 3 to 7 ms according to the manufacturer. The room lights were on because in a dark room, the calibration of the eye tracker was failing on many participants. Consequently, participants saw nothing of the scene except the phosphenes, although part of the room was visible in their peripheral vision.

\begin{figure}[tbp]
  \centering
  \includegraphics[width=1.0\linewidth]{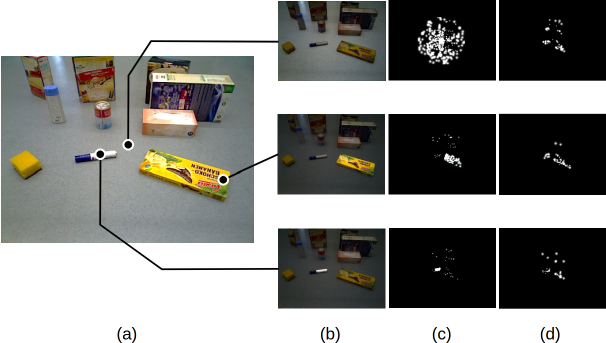}
  \caption{Illustration of the two phosphene conditions at three gaze positions, one per row; the figure shows no experimental data. (a) An example scene from the ARID10 and ARID20 subsets of the Object Clutter Indoor Dataset used in the experiment. The black dots mark the three gaze positions and are drawn for illustration only; no gaze marker was visible to participants during the experiment. (b) The segment selected by gaze-controlled segmentation (GCS) at each gaze position: the floor in the first row, the food box in the second and the marker in the third. (c) Simulated phosphene vision under GCS, in which the selected segment is filled and the edges of the rest of the scene are superimposed at about 30\% of its intensity. (d) Simulated phosphene vision under Edge Detection at the same gaze positions, in which only the edges of the scene are rendered, at full intensity.}
  \label{fig:base}
\end{figure}

In the following part we will give an example of what the two methods look like under the phosphene simulator. Figure \ref{fig:base} shows the segmentation result on a specific point and the phosphene representation under GCS and Edge Detection. For each method we focus the participant's gaze into three spots: the center of the image (background), the food box and the marker to the left of the food box. For GCS, we can see that in the first image the background is highlighted. In the last image we can see that the algorithm highlights part of the marker, making the object very identifiable. In the second image we can see the food box highlighted as well as smaller, slightly dimmer phosphenes representing the edges of the rest of the objects.

In the Edge Detection example, each phosphene keeps a constant position in gaze-centred coordinates, and therefore a constant retinotopic eccentricity, as the participant looks around. For that reason, the participants can never see a peripheral phosphene in the center of their visual field. Characteristically, in the last image, the edges of the marker become finer revealing more details about the object.

\subsection{Analysis}
\label{sec:analysis}

A response was scored as correct when, on target-present trials, the participant clicked inside the requested object's segmentation mask, and on target-absent trials when they correctly reported the object to be absent; clicking on a target-absent trial, or reporting ``absent'' on a target-present trial, was scored incorrect. The best fixed guessing strategy for the present/absent decision alone is to answer present on every trial, which scores 0.71. Because a correct response on a target-present trial requires both the present/absent decision and a correct localisation, the combined measure has no single chance level. On target-present trials, an observer who answered present and clicked a uniformly chosen annotated object would score 0.24, and one who clicked a uniformly chosen pixel 0.02, while answering absent scores 0. Because only a subset of the objects in each scene is annotated, 0.24 is a conservative upper bound on the localisation null; the full set of null strategies is tabulated in the Supplementary Material (Section S2). 

We analysed accuracy with Bayesian generalized linear mixed models (GLMMs) fitted with the Bambi package \cite{Capretto2022} (built on PyMC), using a Bernoulli response and a logit link. We chose mixed models because they analyse the trial-level outcomes directly and account for the repetition of both participants and images, and a Bayesian framework because weakly informative priors keep the full random-effects structure estimable with eight participants and yield the effects of interest directly as posterior contrasts and probabilities. Models included fixed effects for the viewing method and the relevant scene factors (clutter, target presence and object shape) with their interactions, together with by-participant random intercepts and random slopes for method, i.e.\ $(1 + \text{method} \mid \text{participant}) + (1 \mid \text{item})$. No by-item random slope for method was included in the accuracy models, because adding it did not reliably improve leave-one-out predictive accuracy (difference in expected log predictive density 2.0, SE 2.0; Supplementary Material, Section S3); with the slope added, the GCS minus Edge Detection contrast on target-present trials was essentially unchanged, at 9.45 percentage points (95\% HDI [4.62, 14.31]). Priors were weakly informative, with Normal(0, 1.5) priors on the intercept and Normal(0, 1) priors on the fixed effects of the accuracy models on the log-odds scale, HalfNormal(2.6) priors on the participant and item standard deviations, and, for the response-time model, corresponding priors of Normal(0, 3.5) on the intercept, Normal(0, 5) on the fixed effects, and HalfNormal(3.5) on the standard deviations on the log scale, with HalfCauchy(1) on the shape parameter; we drew 2000 posterior samples from each of four chains after 2000 warm-up samples, and all reported models converged ($\hat{R} \leq 1.01$) with bulk and tail effective sample sizes above 1350. We summarise effects as posterior mean contrasts on the probability scale (percentage points), averaged over the observed participants and images, with 95\% highest-density intervals (HDIs) and the posterior probability $P$ that the effect favours GCS. Our primary comparison is restricted to target-present (true) trials, where accuracy reflects finding and identifying the requested object; the full model spanning target-present and target-absent trials is reported in Section~\ref{sec:overall}, and the model-validation checks in the Supplementary Material (Section S3), which also gives the prior specifications in full (Section S4), the convergence diagnostics for every fit (Section S5), and a sensitivity analysis refitting every model under four alternative prior specifications (Section S6). Response times were analysed with an analogous Gamma generalized linear mixed model (log link), $\text{Time} \sim \text{method} + (1 + \text{method} \mid \text{participant}) + (1 + \text{method} \mid \text{item})$, summarised as the posterior mean difference in seconds with its 95\% HDI and the posterior probability that GCS is slower. This model was fitted to both target-present and target-absent trials; target presence was not entered as a predictor, and the normal-vision condition was not included. To check for within-session change, we fitted a logistic GLMM with trial number, method, and their interaction, $\text{Success} \sim \text{trial}\times\text{method} + (1 + \text{method} \mid \text{participant}) + (1 \mid \text{item})$, on target-present trials, and an analogous model with the counterbalanced starting method and its interaction with condition, $\text{Success} \sim \text{method}\times\text{start} + (1 + \text{method} \mid \text{participant}) + (1 \mid \text{item})$.

We distinguish two classes of analysis throughout. The method contrast is a within-participant comparison estimated from 608 trials per condition with each participant serving as their own control, and is the analysis the design was built to support. Analyses that depend on variation between participants---condition order and the questionnaire responses---and analyses that subdivide trials by scene factors---object shape, clutter level and trial number---are supported by correspondingly less data. We report these as exploratory throughout, without correction for multiplicity and without treating them as confirmatory tests.

Because the analysis is Bayesian, we characterise the resolution of the design by the precision of the posterior rather than by a power calculation. Under the primary model the GCS minus Edge Detection contrast on target-present trials is estimated with a 95\% HDI of width 9.9 percentage points, so the design separates effects of roughly five percentage points and larger from zero, while smaller differences remain unresolved. Interval widths for every reported contrast are given in the Supplementary Material (Section S7). Throughout, we describe an effect as credible when its 95\% HDI excludes zero, and we report the posterior probability that the effect lies in the stated direction alongside the interval.

To characterise gaze, we computed object-aware metrics from the eye-tracking samples and the scene masks: the fraction of gaze on objects versus the background, the number of distinct objects inspected, whether and when gaze first reached the target region (within a $30$~px tolerance, which is 0.95 degrees of visual angle at our viewing geometry; the Supplementary Material (Section S8) repeats these analyses at tolerances of 0, 15, 30 and 45 px), and---on error trials---whether the non-target object that attracted the most gaze shared the target's shape. Because the two conditions were logged at different frame rates ($\approx\!30$ vs.\ $\approx\!40$~Hz---the rendering rate rather than the tracker's), count- and velocity-based measures are confounded, so we relied on rate-invariant, fraction-of-time gaze measures.

\section{Results}

\subsection{Overall task performance}
\label{sec:overall}

The requested object was present in the scene on roughly $70\%$ of trials and absent on the remainder; a response was correct when the participant identified the target or, on target-absent trials, reported it to be absent (Section~\ref{sec:analysis}). The normal-vision condition was included to establish a performance baseline: accuracy on the same task was near ceiling (mean accuracy $0.96$ on target-present trials, Table~\ref{tab:accuracy}, and 0.99 on target-absent trials, with a mean response time of 1.6 s).

The normal-vision condition was always administered last and without a mid-session break. It is not a condition in the method comparison but a reference for its scale as under normal vision participants erred roughly fifteen times less often and responded five to seven times faster. Practice and fatigue act in opposite directions and, by our within-session estimates, amount to a few percent per ten trials, so neither can account for differences of this size; we therefore report this condition descriptively and did not counterbalance it.

\begin{figure}[tbp]
  \centering
  \includegraphics[width=1.0\linewidth]{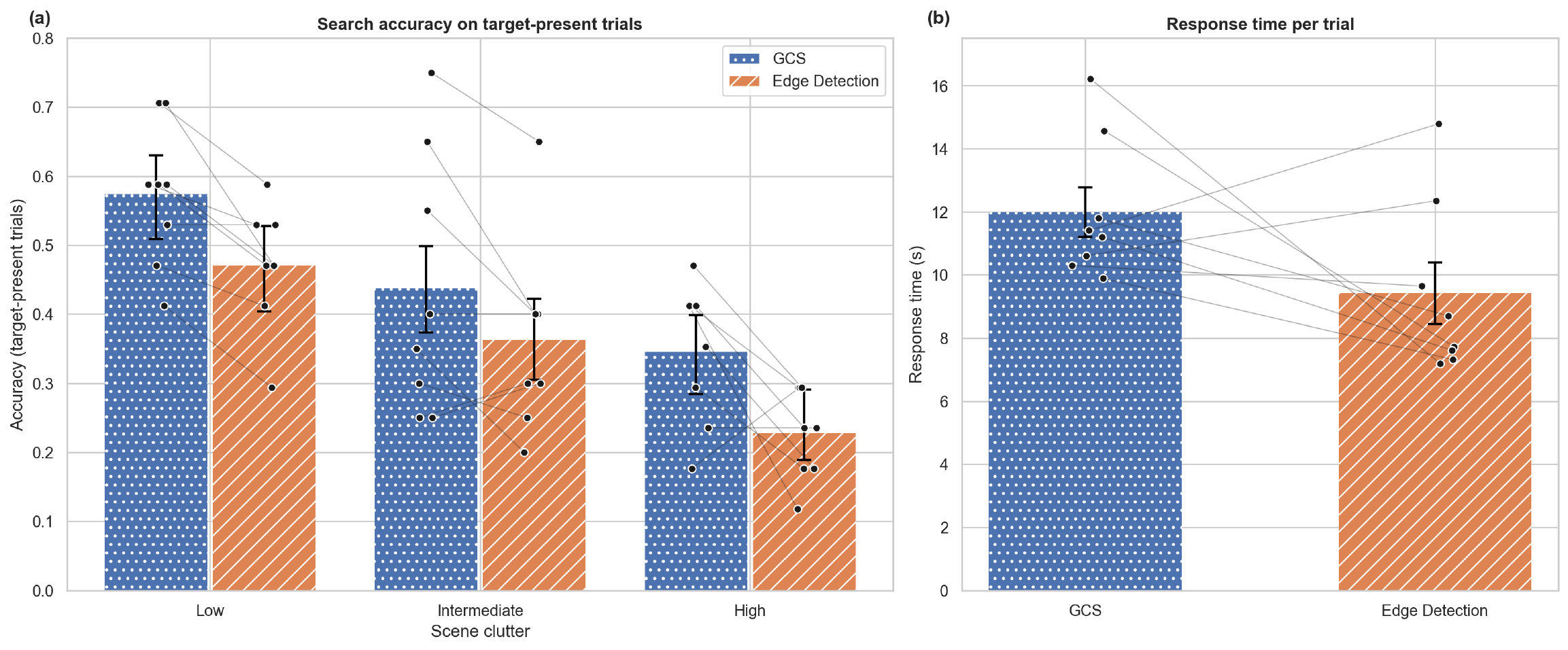}
  \caption[Accuracy and response time by method]{Accuracy and response time under gaze-controlled segmentation (GCS) and Edge Detection, for the eight analysed participants. (a) Accuracy on target-present trials by scene clutter (low, 1 to 3 objects; intermediate, 5 to 8; high, 9 to 12). Bars show the proportion of correct trials averaged over participants, black points each participant's proportion, and lines join each participant's two methods. Error bars are 95\% highest-density intervals (HDIs) from a Bayesian logistic mixed-effects model with method, clutter and their interaction as fixed effects and participants and images as crossed random effects, averaged over the observed participants and images. GCS was more accurate at every clutter level, by 10.3, 7.7 and 10.5 percentage points at low, intermediate and high clutter; the HDI of the difference excludes zero at low and high clutter but not at intermediate clutter. (b) Response time per trial over all trials, target-present and target-absent. Bars show the mean over participants, black points each participant's mean, and error bars $\pm 1$ standard error of the mean across participants, a descriptive summary rather than a model interval. A Bayesian Gamma mixed-effects model with method as a fixed effect and participants and images as crossed random effects estimated GCS to be slower by 2.7 s (95\% HDI [-0.8, +6.6] s; posterior probability 0.95 that GCS is slower); because this HDI includes zero, the difference is not credible.}
  \label{fig:accuracy_true}
\end{figure}

\begin{figure}[tbp]
  \centering
  \includegraphics[width=1.0\linewidth]{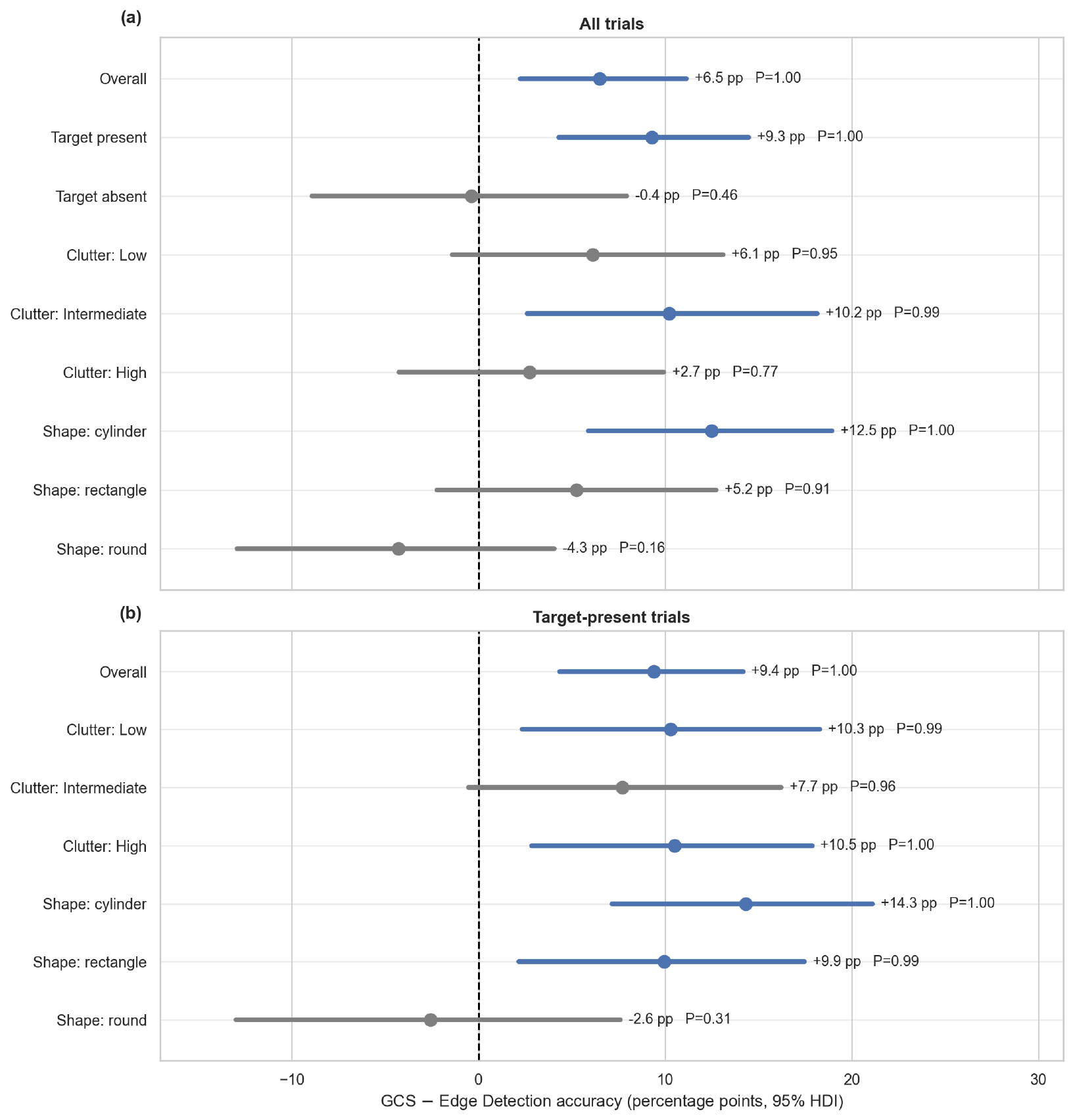}
  \caption[GCS versus Edge Detection accuracy differences]{Difference in accuracy between gaze-controlled segmentation (GCS) and Edge Detection, as GCS minus Edge Detection in percentage points (pp), for the eight analysed participants, (a) across all trials and (b) on target-present trials only. Each row is one contrast: the point is the posterior mean, the horizontal line its 95\% highest-density interval (HDI), and the printed P the posterior probability that the difference favours GCS. Contrasts come from Bayesian logistic mixed-effects models with method and, as the row requires, target presence, clutter or object shape and its interaction with method as fixed effects, and participants and images as crossed random effects; they are averaged over the observed participants and images. A difference is credible when its HDI excludes zero; blue marks a credible difference and grey one whose HDI includes zero. The advantage is carried by target-present trials (+9.3 pp in (a), +9.4 pp in (b)) and is absent on target-absent trials (-0.4 pp). On target-present trials it is positive at every clutter level (credible at low and high clutter) and for cylindrical and rectangular objects, and absent for round objects (-2.6 pp).}
  \label{fig:forest}
\end{figure}

Across all trials, participants were more successful with GCS than with Edge Detection: a participant-level Bayesian GLMM spanning both trial types (Section~\ref{sec:analysis}) estimated the advantage at $+6.5$ percentage points (pp), 95\% highest-density interval (HDI) $[+2.2, +11.1]$, with posterior probability $P=1.00$ that the effect favours GCS (Table~\ref{tab:full}, Figure~\ref{fig:forest}a). This advantage was carried entirely by target-present trials, on which participants identified the target more often with GCS (mean accuracy $0.45$ vs.\ $0.35$; $+9.3$~pp, $[+4.3, +14.5]$, $P=1.00$); on target-absent (detection) trials the two methods did not differ ($-0.4$~pp, $[-8.9, +7.9]$, $P=0.46$). GCS thus helps participants identify an object that is present, but does not change how reliably they can tell that a requested object is absent. This advantage is an effect of the integrated GCS representation rather than of the segmentation step alone, since the two methods also differ in whether the selected region is filled, in the stimulation current, and therefore the size and brightness, of the background edges, and in the resulting number and distribution of activated phosphenes. Both phosphene conditions exceeded the random-object null of 0.24 for this measure. Figure~\ref{fig:accuracy_true}a and Table~\ref{tab:accuracy} break accuracy on target-present trials down by scene clutter.

Because each session contained 54 target-present and 22 target-absent trials, accuracy would also rise if participants simply answered present more readily under GCS, without perceiving the scene any better. A signal detection analysis of the present/absent decision, reported in full in the Supplementary Material (Section S9), separates this readiness to respond from the ability to tell the two kinds of trial apart. GCS raised the hit rate by 8.2 pp (95\% HDI [+3.0, +13.7]) with no credible change in the false-alarm rate (+1.9 pp, [-7.0, +11.2]). The resulting increase in sensitivity, 0.19 in d' ([-0.09, +0.47], P = 0.91), was positive but not credible, and the shift towards answering present lay on the boundary of credibility (criterion change -0.14, [-0.28, 0.00]). A readier response alone cannot explain the accuracy advantage on target-present trials, because it would turn some absent responses into clicks on the wrong object as well as into clicks on the target. Wrong-object clicks did not change (-1.1 pp, [-6.2, +4.0]), whereas absent responses fell by 8.3 pp ([-13.4, -3.0]) as correct identifications rose, so the advantage reflects more targets being identified and not merely more responses being made.

Participants took longer with GCS than with Edge Detection (model-based means $11.5$ vs.\ $8.8$~s; Figure~\ref{fig:accuracy_true}b). A participant-level Gamma generalized linear mixed model (Section~\ref{sec:analysis}) estimated the difference at $+2.7$~s (95\% HDI $[-0.8, +6.6]$; equivalently a ratio of $1.31\times$, $[0.92, 1.75]$), with posterior probability $P=0.95$ that GCS is slower. The accuracy advantage is credible on both all trials and target-present trials ($P=1.00$ and $1.00$), so the evidence that GCS is slower is weak, since the 95\% HDI narrowly includes zero. Most participants were slower with GCS, and two were faster, so the effect varies across participants but the bulk of the posterior favours a real slowdown. Both methods required well over eight seconds per trial, underlining the difficulty of object identification under simulated phosphene vision. We do not read this difference as a speed--accuracy trade-off as response time was not resolved into components, so how the additional time was spent---for example, in verifying a highlighted region---cannot be determined here.

The speed cost did not change across the session. In a model of response time with method, trial number and their interaction, both methods became faster as the session progressed---Edge Detection by 4.6\% per ten trials (95\% HDI [-7.4, -1.9]) and GCS by 2.8\% ([-6.3, +0.9])---and the method-by-trial-number interaction was not credible (+0.019 log-seconds per ten trials, [-0.029, +0.066]). The GCS-Edge difference was +2.3 s ([-0.1, +4.7]) at the start of a session and +3.1 s ([+1.3, +5.0]) at the end, a change of +0.8 s ([-3.2, +4.9]). The additional time GCS requires is therefore a stable cost rather than a sign of adaptation or fatigue.

\begin{table}[tbp]
  \centering
  \caption{Accuracy on target-present trials by viewing condition and scene clutter (low, 1 to 3 objects; intermediate, 5 to 8; high, 9 to 12), for the eight analysed participants. Each value is the proportion of trials answered correctly, averaged over participants, who all saw the same trials; the table is descriptive and shows no uncertainty. GCS, gaze-controlled segmentation. Normal vision, in which the scenes were shown without phosphene simulation, is a near-ceiling reference and did not enter the method comparison. Model-based differences between GCS and Edge Detection, with their 95\% highest-density intervals, are given in Table~\ref{tab:full} and Figure~\ref{fig:forest}.}
  \label{tab:accuracy}
  \begin{tabular}{lcccc}
    \toprule
    Method & Overall & Low & Intermediate & High \\
    \midrule
    Normal vision & 0.96 & 0.93 & 0.97 & 0.98 \\
    GCS & 0.45 & 0.57 & 0.44 & 0.35 \\
    Edge Detection & 0.35 & 0.47 & 0.36 & 0.23 \\
    \bottomrule
  \end{tabular}
\end{table}

\begin{table}[tbp]
  \centering
  \caption{Difference in accuracy between gaze-controlled segmentation (GCS) and Edge Detection, as GCS minus Edge Detection in percentage points (pp), across all trials of the eight analysed participants. Each difference is the posterior mean from a Bayesian logistic mixed-effects model with participants and images as crossed random effects, averaged over the observed participants and images. The overall and target-presence rows come from one model with method, target presence and their interaction as fixed effects; the clutter and shape rows come from models with a method-by-clutter or method-by-shape interaction and are averaged over target-present and target-absent trials. HDI, highest-density interval; $P(\text{GCS}>\text{Edges})$, the posterior probability that GCS is more accurate than Edge Detection. A difference is credible when its 95\% HDI excludes zero. The advantage is confined to target-present trials and is absent on target-absent trials.}
  \label{tab:full}
  \begin{tabular}{lccc}
    \toprule
    Condition & Difference (pp) & 95\% HDI & $P(\text{GCS}>\text{Edges})$ \\
    \midrule
    Overall (all trials) & $+6.5$ & $[+2.2, +11.1]$ & $1.00$ \\
    Target present & $+9.3$ & $[+4.3, +14.5]$ & $1.00$ \\
    Target absent & $-0.4$ & $[-8.9, +7.9]$ & $0.46$ \\
    Clutter: Low & $+6.1$ & $[-1.4, +13.1]$ & $0.95$ \\
    Clutter: Intermediate & $+10.2$ & $[+2.6, +18.1]$ & $1.00$ \\
    Clutter: High & $+2.7$ & $[-4.3, +9.9]$ & $0.77$ \\
    Shape: cylinder & $+12.5$ & $[+5.9, +18.9]$ & $1.00$ \\
    Shape: rectangle & $+5.2$ & $[-2.2, +12.7]$ & $0.91$ \\
    Shape: round & $-4.3$ & $[-12.9, +4.1]$ & $0.16$ \\
    \bottomrule
  \end{tabular}
\end{table}

\subsection{When does gaze control help?}

Our primary comparison (Section~\ref{sec:analysis}) is restricted to target-present trials, where accuracy reflects finding and identifying the requested object. Figure~\ref{fig:forest}b summarises the GCS$-$Edge Detection accuracy difference from this model, overall and within conditions. The overall advantage of $+9.4$~pp (95\% HDI $[+4.3, +14.2]$, $P=1.00$) agrees closely with the full-model estimate above and was present at every level of scene clutter---$+10.3$~pp at low ($[+2.3, +18.3]$, $P=0.99$), $+7.7$~pp at intermediate ($[-0.5, +16.2]$, $P=0.97$) and $+10.5$~pp at high clutter ($[+2.8, +17.9]$, $P=1.00$); the benefit is not confined to, or growing with, any particular clutter level. The advantage did, however, depend on object shape. It was strongest for cylindrical objects ($+14.3$~pp, $[+7.2, +21.1]$, $P=1.00$), also positive for rectangular objects ($+10.0$~pp, $[+2.1, +17.5]$, $P=1.00$), and absent---if anything slightly reversed---for round objects ($-2.6$~pp, $[-13.0, +7.6]$, $P=0.31$). Absolute accuracy fell with clutter for both methods, as expected. The corresponding clutter- and shape-specific contrasts from the all-trials model (Table~\ref{tab:full}) are averaged over present and absent trials and so differ slightly in magnitude, but tell the same story, the advantage is largest for cylindrical objects and weakest, or absent, for round ones.

These within-condition breakdowns subdivide the trials of eight participants and were not planned as confirmatory tests. We report them as exploratory and read the shape dependence as a hypothesis for future work rather than as an established moderator.

\subsection{Gaze behaviour}
\label{subsec:gaze}

\begin{figure}[tbp]
  \centering
  \includegraphics[width=0.62\linewidth]{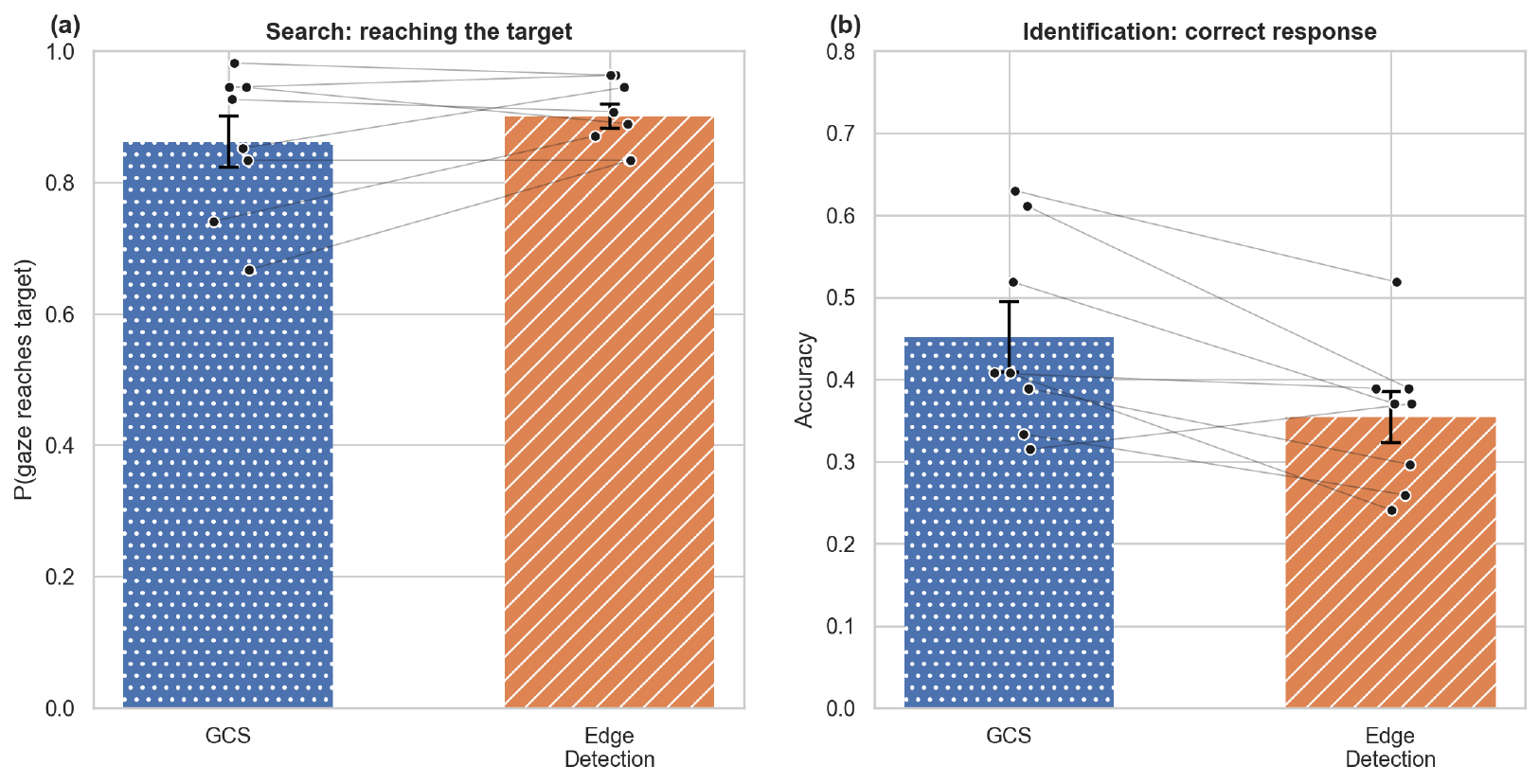}
  \caption[Search versus identification]{Reaching the target and identifying it on target-present trials under gaze-controlled segmentation (GCS) and Edge Detection, for the eight analysed participants. (a) Proportion of trials on which gaze came within 30 pixels (0.95 degrees of visual angle) of the target's segmentation mask, which did not differ reliably between the methods (Edge Detection 0.90, GCS 0.86; two-sided paired t-test on the participant means, p = .21). (b) Proportion of trials answered correctly, which was higher under GCS (GCS 0.45, Edge Detection 0.35; difference +9.4 percentage points, 95\% highest-density interval [+4.3, +14.2], from a Bayesian logistic mixed-effects model with participants and images as crossed random effects). Bars show means across participants, black points each participant's mean, and lines join each participant's two methods; error bars are $\pm 1$ standard error of the mean across participants, a descriptive summary. Gaze reached the target about as often under both methods, yet identification was less accurate under Edge Detection, a pattern consistent with the advantage arising at the identification stage, as the timing of the highlight would predict.}
  \label{fig:reach_identify}
\end{figure}

To ask whether the GCS advantage arises while \emph{searching} for the target or while \emph{identifying} it once found, we examined the gaze record. Gaze reached the target region on as many target-present trials under Edge Detection as under GCS---if anything slightly more often ($0.90$ vs.\ $0.86$ within a $30$~px, that is 0.95 degrees of visual angle, tolerance)---and reached it at the same point in the trial, yet Edge Detection was the \emph{less} accurate method. Participants could therefore bring their gaze to the target at least as readily with edges and still fail to identify it, a pattern consistent with the advantage arising at the identification stage rather than in visual search (Figure~\ref{fig:reach_identify}).

Beyond this, gaze behaviour did not reveal a distinct process behind the advantage (Table~\ref{tab:gaze}). Once trials were separated by outcome, GCS and Edge Detection did not differ in how gaze approached the target (time to first contact), how long it remained once there, how many objects were inspected, or how gaze was divided between objects and the background. The one systematic spatial difference---under GCS gaze fell more on object \emph{bodies}, and under Edge Detection more on object \emph{boundaries}---follows directly from what each method renders (a filled interior versus a sparse outline) and did not itself predict success. The greater total time GCS users appeared to spend on the target ($0.21$ vs.\ $0.18$ of the trial) was fully accounted for by GCS's higher proportion of correct trials. Within correct trials, and within incorrect trials, the two methods were indistinguishable, so it reflects the accuracy difference rather than a separate viewing strategy. Measures of exploration \emph{dynamics}---scan speed, coverage, transition counts---could not be compared directly, because the two conditions were logged at different frame rates ($\approx\!30$ vs.\ $\approx\!40$~Hz) and any count- or velocity-based statistic changes with the sampling rate; the measures that do not depend on the frame rate showed no pattern specific to GCS.

\begin{table}[tbp]
  \centering
  \caption{Gaze measures under gaze-controlled segmentation (GCS) and Edge Detection (Edge Det.) for the eight analysed participants, on target-present trials except where a row states otherwise. The target was reached when gaze came within 30 pixels (px; 0.95 degrees of visual angle) of its segmentation mask; time to first contact and dwell are fractions of the trial duration, with dwell measured at a strict 0 px tolerance; shape-matched confusion is the proportion of error trials on which the non-target object that attracted most gaze resembled the target in shape. Each row gives the mean of the eight participants' per-participant means for each method, and the p-value of a two-sided paired t-test on those per-participant means (degrees of freedom, df = 7), so that trials are not treated as independent. The seven comparisons are descriptive checks rather than confirmatory tests and no correction for multiple comparisons was applied; none is significant at any conventional threshold (smallest p = .21), and a correction could only raise these values. No measure favoured GCS; reaching the target if anything favoured Edge Detection, and the greater whole-trial dwell on the target under GCS disappeared once trials were split by outcome.}
  \label{tab:gaze}
  \begin{tabular}{lccc}
    \toprule
    Measure & GCS & Edge Det. & $p$ \\
    \midrule
    Reached target (within 30\,px, 0.95$^\circ$) & 0.86 & 0.90 & .21 \\
    Time to first contact, correct trials (fraction of trial) & 0.23 & 0.23 & .57 \\
    Retention on target (correct trials) & 0.73 & 0.76 & .48 \\
    Distinct objects inspected (all trials) & 3.8 & 3.7 & .85 \\
    Dwell on target, correct trials & 0.37 & 0.35 & .64 \\
    Dwell on target, incorrect trials & 0.09 & 0.09 & .77 \\
    Shape-matched confusion (error trials) & 0.27 & 0.27 & .79 \\
    \bottomrule
  \end{tabular}
\end{table}

On error trials, the non-target object that attracted the most gaze resembled the target in shape equally often under the two methods ($27\%$ of error trials in both; Table~\ref{tab:gaze}), so Edge Detection did not induce more shape confusions than GCS.

Taken together, these analyses indicate that the GCS advantage is perceptual---a filled object is easier to identify than a sparse outline---and that, at the temporal resolution of our recording, this benefit does not surface as a measurable difference in where or how participants looked.

\subsection{Learning and order effects}
\label{sec:learning}

\begin{figure}[tbp]
  \centering
  \includegraphics[width=1.0\linewidth]{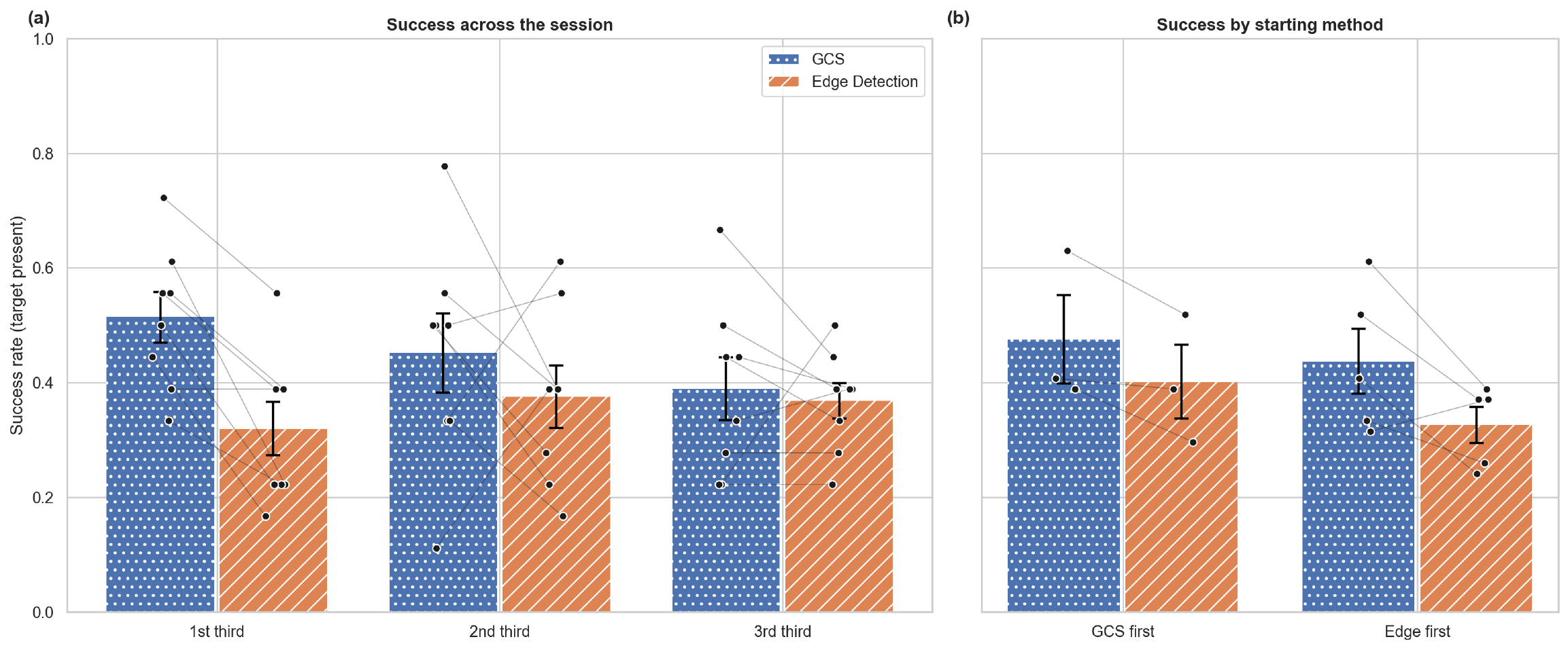}
  \caption[Accuracy across the session and by starting method]{Accuracy on target-present trials across the session and by starting method, under gaze-controlled segmentation (GCS) and Edge Detection, for the eight analysed participants; both analyses are exploratory. (a) Accuracy in the first, second and third thirds of each session, by trial position. (b) Accuracy of the three participants who started with GCS and the five who started with Edge Detection. Bars show the proportion of correct trials, black points each participant's proportion, and lines join each participant's two methods; error bars are $\pm 1$ standard error of the mean across participants, a descriptive summary. Estimates come from Bayesian logistic mixed-effects models with method and trial number (a), or method and starting method (b), and their interaction as fixed effects, and participants and images as crossed random effects; intervals are 95\% highest-density intervals (HDIs). In (a) the GCS advantage is largest in the first third and narrows over the session (method-by-trial-number interaction -0.18 log-odds per 10 trials, 95\% HDI [-0.38, -0.00]), but trial position is confounded with image identity, because every participant saw the images in the same order. In (b), neither the starting method (+0.39 log-odds, 95\% HDI [-0.61, +1.39]) nor its interaction with method (-0.12 log-odds, [-1.06, +0.83]) was credible.}
  \label{fig:learning}
\end{figure}

We asked whether performance drifted over the course of a session. Pooled across methods, success on target-present trials did not change reliably with trial number ($+0.11$ log-odds per 10 trials, 95\% HDI $[-0.04, +0.26]$), indicating no general learning or fatigue effect on accuracy. There was a trial-number\,$\times$\,method interaction ($-0.18$ log-odds per 10 trials, $[-0.38, -0.00]$; posterior probability $0.98$ that it is negative). Because a single presentation order was used for all participants, trial position is perfectly confounded with image identity. Any tendency for later items to be easier or harder is indistinguishable from a change in performance over the session. We therefore report this analysis as exploratory and do not interpret it as evidence of learning or fatigue (Figure~\ref{fig:learning}a). The counterbalanced starting method is examined next.

Because the GCS and Edge Detection sessions were counterbalanced---the first four recruited participants received GCS first and the remainder Edge Detection first (subjects $1$, $3$ and $4$ versus $5$--$9$ among those analysed)---we also checked whether the starting method affected success. The result was negative and neither the starting method (main effect $+0.39$ log-odds, 95\% HDI $[-0.61, +1.39]$, $P=0.80$) nor its interaction with condition ($-0.12$ log-odds, $[-1.06, +0.83]$, $P=0.39$) had a credible effect on target-present accuracy (Figure~\ref{fig:learning}b). With only three GCS-first and five Edge-first participants these estimates are necessarily imprecise, but they give no indication that the counterbalancing biased the comparison.

\subsection{User preference}

\begin{figure}[tbp]
  \centering
  \includegraphics[width=0.7\linewidth]{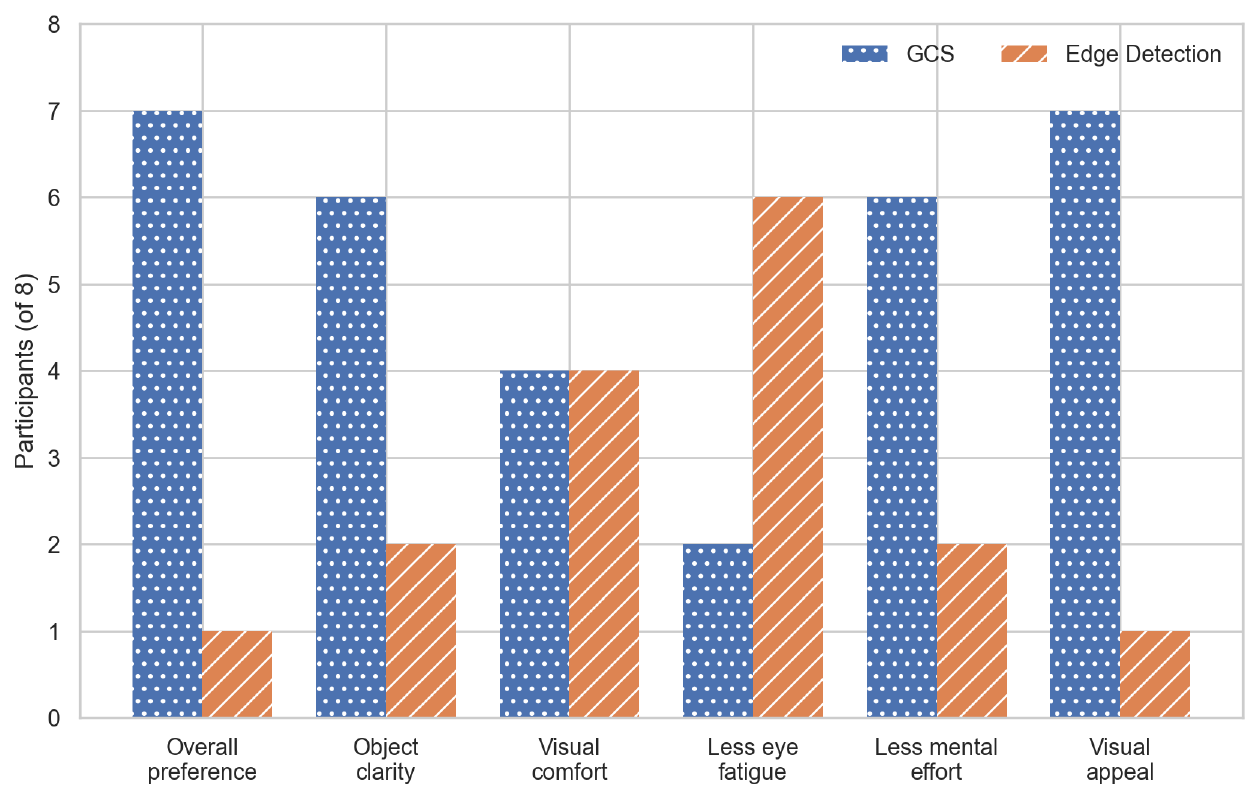}
  \caption[Preference between GCS and Edge Detection]{Preferences of the eight analysed participants between gaze-controlled segmentation (GCS) and Edge Detection on six criteria of the post-experiment questionnaire: overall preference, clarity of object appearance, visual comfort, reduction of eye fatigue, reduction of mental effort and visual appeal. Bars show how many participants chose each method; each participant made one choice per criterion, so the bars show every observation. GCS was preferred overall (7 of 8) and for clarity (6 of 8), mental effort (6 of 8) and visual appeal (7 of 8), Edge Detection for less eye fatigue (6 of 8), and visual comfort was split (4 of 8). Two-sided binomial tests against an even split, without correction for the six criteria, give p = .07 for 7 of 8 and p = .29 for 6 of 8, so these are descriptive tendencies.}
  \label{fig:experience questionnaire}
\end{figure}

After the experiment, the eight analysed participants compared the two methods on a short questionnaire (Figure~\ref{fig:experience questionnaire}). A majority favoured GCS for overall preference (7/8), clarity of object appearance (6/8), reduced mental effort (6/8) and visual appeal (7/8); opinions on visual comfort were split (4/8); and Edge Detection was preferred for causing less eye fatigue (6/8). With only eight participants these proportions are individually consistent with chance (two-sided binomial tests: $7/8$, $p=.07$; $6/8$, $p=.29$), so we report them as descriptive tendencies rather than reliable effects. When asked whether each method became easier to use over the session, $6/8$ participants reported improvement with GCS versus $2/8$ with Edge Detection.

\section{Discussion}

In this study, we aimed to design a phosphene vision algorithm in which the object that is filled in the phosphene representation is selected by the user's gaze. To test the capabilities of our algorithm, we designed an object search-and-localization experiment in which the participants were subjected to simulated phosphene vision. Using an eye-tracker, participants were able to highlight objects by looking at them on a screen. We compared our method with a widely used method in artificial vision research, edge detection, as well as healthy vision. In the following sections we interpret the results from the comparison between the different presentation methods.

\subsection{How gaze control helps}

On target-present trials, GCS produced a modest but credible improvement in identification accuracy over Edge Detection (about $+9$ percentage points). The gaze record suggest the improvement happens at the identification stage. Participants were no more likely to bring their gaze to the target under GCS, yet identified it more often, so the benefit appears to lie in identifying the object rather than in finding it, as the timing of the highlight would predict. Filling the gazed object's silhouette helps users recognise what they are already looking at, and this benefit did not surface as any measurable difference in gaze behaviour at the resolution we recorded. This specificity also holds across trial types. When the requested object was absent, the two methods performed alike, since there is no object for a filled highlight to clarify. GCS therefore aids the identification of a target that is present rather than the decision that a target is absent. We describe this as a property of the GCS representation as a whole. Which of its components carries the effect (object-level selection, the filled rendering, or the increased phosphene drive within the selected region) is not identified by this experiment.

The benefit depended on object shape as it was strongest for cylindrical objects, also present for rectangular ones, and absent---if anything slightly reversed---for round objects, consistent with the highlight helping most when an object's silhouette is otherwise ambiguous. Across scene clutter it was more uniform, remaining present at every level rather than being confined to one.

Beyond the object/background split, the gaze record offered no evidence for a distinct viewing strategy behind the advantage. The two methods did not differ in how participants divided their gaze between objects and the background. The one systematic spatial difference---gaze falling more on object bodies under GCS and more on object boundaries under Edge Detection---follows directly from what each method renders and did not itself predict success. The greater whole-trial dwell on the target under GCS was fully accounted for by its higher proportion of correct trials, disappearing once trials were split by outcome, and on error trials the most-gazed non-target object resembled the target in shape equally often under both methods ($27\%$). Edge Detection therefore did not induce more shape confusions, and the advantage does not reflect a different way of exploring the scene.

\subsection{The cost in time}

The identification gain came at a cost in time. GCS was slower than Edge Detection by about $2.7$~s per trial ($11.5$ vs.\ $8.8$~s, a $1.31\times$ ratio), with posterior probability $0.95$ that GCS is the slower method. Because the accuracy advantage is credible under the crossed model ($P=1.00$--$1.00$), we treat it as the primary result; the time cost is reported descriptively, with no claim that time was traded for accuracy. The slowdown varied across participants---most were slower under GCS, though two were faster---but the bulk of the posterior favours a real difference. Both methods nonetheless required well over eight seconds per trial, underlining how demanding object identification remains under simulated phosphene vision.

\subsection{Stability over a session and user preference}

The GCS advantage was not perfectly stable within a session. Pooled across methods, accuracy showed no general learning or fatigue over trials, but there was a trial-number\,$\times$\,method interaction that we report as exploratory. Because a single presentation order was used for all participants, trial position and image identity are the same variable, so this coefficient cannot distinguish a change in performance over the session from a difference between early and late images. We therefore do not interpret it, and longer training and multi-session designs would be needed to settle the question. Reassuringly, the counterbalanced starting method had no credible effect on accuracy, so the comparison does not appear to have been biased by order, although with only three GCS-first and five Edge-first participants this check is necessarily imprecise.

Participants' own preferences tended to favour GCS: a majority chose it for overall preference, clarity of object appearance, reduction of mental effort and visual appeal; opinions on visual comfort were split; and Edge Detection was preferred only for causing less eye fatigue. More participants also felt that GCS became easier to use over the session than said the same of Edge Detection. With only eight participants these proportions are individually consistent with chance, so we treat them as descriptive tendencies rather than reliable effects; nonetheless they align with the direction of the accuracy results and suggest that the eye fatigue of sustained gaze control---rather than the representation itself---is the main experiential cost of the method.

\subsection{Limitations and future work}

While this study provides valuable insights into the effectiveness of a gaze-controlled segmentation method, several limitations must be acknowledged. First, due to computational limitations, the experiment was conducted using pre-calculated masks. This does not constitute a real-time solution and the results should only be considered as an indication of the capabilities of such an approach. As the results demonstrated a promising performance by the GCS method, optimizing the segmentation algorithm to process the images in real-time should be a primary focus of future work. Real-time variants of SAM optimized for speed, such as MobileSAM \cite{MobileSAM} and FastSAM \cite{FastSAM}, already show promise by prioritizing performance at the cost of some robustness, and could stand in for the full model in a live gaze-prompted pipeline.

A second aspect of our experiment that was impacted by computational limitations was the task choice. Due to the speed at which we could run SAM at the time of the experiment design, it was not possible for us to choose a design that could test a participant's navigational skills. Since then, we were able to improve our hardware setup and our computational pipeline, which we will be utilizing to delve further into how to implement and test such methods in more naturalistic settings. A natural next step is an immersive VR experiment in which participants can move through a realistic outdoor environment.

Another shortcoming that needs to be taken into consideration is that while our experiment allows eye movements to alter how the scene is rendered, the camera viewpoint was fixed, so head and body movement did not affect the scene. The participants had to use a chin rest and were instructed not to move their heads, which admittedly makes the whole experiment a more unpleasant experience than it should be, possibly influencing the results. Also importantly, in the real world, head movements are utilized to allow for more degrees of freedom and perspective, which in the experimental setup would translate into better object recognition performance as well as a more natural experience for the participants. Ideally this experimental setup can be replaced by using a VR headset with eye-tracking capabilities.

Finally, after the experiment sessions, many participants expressed their lack of confidence when selecting objects particularly commonly with the Edge Detection method. A controlled way for the participants to express their confidence levels about the choices was thus identified as an important metric to include in future studies of the kind. Including questions regarding confidence in our questionnaire, or by directly asking participants about their confidence levels in their choices after each trial would let us gain insights about this important part of user experience.

A further consideration concerns the relationship between the two conditions, which bears directly on our hypothesis that they would elicit distinct exploration strategies. In the GCS condition, the gaze-selected object highlight is superimposed on the edge image at reduced intensity, so the GCS representation contains the edge information together with the highlighted object region. The two conditions are therefore not independent: they share the same underlying edges and differ only by the added highlight. Because both expose participants to the same edge information for guiding their gaze, the scope for the two methods to drive genuinely different exploration is limited, and the absence of a clear difference in gaze behaviour may reflect this overlap rather than a true equivalence of the two representations. For the same reason, our comparison evaluates the integrated gaze-driven method against edge detection and does not isolate the contribution of the segmentation itself from the added object highlight or from the interactivity. A more decisive test would add two conditions. In the first, a filled patch of fixed shape, such as a disc, would follow the participant's gaze in place of the segmented object; its area would be matched on each trial to that of the object, and it would be shown on the same 30\% edge background, so that a difference from GCS would reflect the shape of the object alone. In the second, only the 30\% edges would be shown, so that comparing it with the patch condition would reveal what filling the region under gaze adds. Both comparisons would require reporting the number of active electrodes and the summed phosphene intensity per trial, because matching the area in the image does not guarantee that the same number of phosphenes is activated.

A limitation specific to gaze-contingent methods concerns our participants: all were normally sighted and viewed a simulation, so their behaviour does not map straightforwardly onto that of implant users. Long-term blindness alters oculomotor control, including reduced fixation stability and nystagmus, and a method whose benefit is delivered entirely through the gaze point is more exposed to this than a gaze-independent rendering. Our results therefore establish an effect within simulated phosphene vision in sighted observers, and motivate rather than predict a benefit in implant users. Evaluating GCS with gaze data from blind participants is a necessary next step.

The phosphenes themselves also follow the assumptions of the simulator rather than the percepts of any implant user. In particular, because of the implementation error described in Section 2.6, making the phosphenes as bright as those participants saw required a current of $600\mu A$ in the simulator, which is about nine times the average current at which implant users detected phosphenes \cite{Fernndez2021} and would not be used in an implant. At clinical currents a similar brightness would be reached within a few frames of continuous stimulation, but the phosphenes would be about half as wide. Because both methods were rendered with the same settings, the comparison between them is unaffected, but larger, overlapping phosphenes may make a filled object easier to see than outlines, so the GCS advantage may be smaller with the phosphenes produced at clinical currents, which remains to be tested. Every electrode was modelled with the same response, and phosphene brightness followed a curve fitted to a single implant recipient, whereas in implant users the same stimulation produces phosphenes that differ in size, brightness, color and shape between electrodes and between people. The difference in brightness between the filled object and the reduced-brightness edges, which GCS relies on, may therefore be different in an implant.

A second limitation of scale is electrode count: we simulated 600, close to the number estimated as necessary for functional vision \cite{Cha1992}, whereas recent human trials have implanted between 44 and 378 electrodes \cite{Fernndez2021}\cite{Allen2025}\cite{Holz2026}. Because the GCS benefit derives from filling an object's interior, and the phosphenes falling inside a typical object scale with electrode count, it would be expected to shrink at present-day counts, where a filled mask and an outline may activate a similar, very small set of phosphenes. How the effect scales with electrode count is another direction for future research.

Lastly, current visual neuroprosthesis trials present only a limited number of phosphenes at a time \cite{Fernndez2021} and impose strict safety limits on the stimulation budget. The current we simulated is itself far above those used in clinical practice. Edge detection is favoured partly because it activates few electrodes, so a filled mask is a real safety concern for shape-filling methods like ours. Gaze-controlled selection at least spends that budget on the attended object, but the safety of this approach needs to be evaluated in future research. Presenting the selected object sequentially rather than all at once, as in raster stimulation \cite{Wang2025}\cite{Kasowski2025raster}, may help reconcile shape filling with these limits.

\section{Conclusion}

We developed a gaze-controlled segmentation method in which the point of gaze serves as the prompt that selects the object to be filled in, so that where a user looks determines which object is highlighted.

In a simulated phosphene-vision search task, this approach produced a small, task-dependent improvement in identification accuracy over conventional edge detection. A participant-level analysis suggests that the benefit comes specifically from helping users recognise an object once their gaze has reached it---rather than from helping them find it, or from a distinct exploration strategy---and that it is most pronounced for objects whose silhouette is otherwise ambiguous. While modest, and obtained with phosphenes larger than current implants would produce, these results highlight the potential of gaze-driven, interactive scene representation for neuroprosthetic vision, and motivate future work that isolates the contribution of segmentation, which the present design does not, and that evaluates the approach in more naturalistic, real-time settings.

\roles{\noindent Eleftherios Papadopoulos\par\noindent Conceptualization (lead), Data curation (lead), Formal analysis (lead), Investigation (lead), Methodology (lead), Software (lead), Validation (lead), Visualization (lead), Writing -- original draft (lead), Writing -- review \& editing (lead)

\medskip\noindent Umut G\"{u}\c{c}l\"{u}\par\noindent Conceptualization (supporting), Data curation (supporting), Formal analysis (equal), Investigation (supporting), Methodology (supporting), Software (equal), Supervision (supporting), Validation (supporting), Visualization (supporting), Writing -- original draft (supporting), Writing -- review \& editing (supporting)

\medskip\noindent Ya\u{g}mur G\"{u}\c{c}l\"{u}t\"{u}rk\par\noindent Conceptualization (equal), Data curation (supporting), Formal analysis (equal), Investigation (supporting), Methodology (equal), Software (equal), Supervision (equal), Validation (supporting), Visualization (supporting), Writing -- original draft (equal), Writing -- review \& editing (equal)}

\data{The trial-level behavioural data, gaze recordings and questionnaire responses of all nine recorded participants, including the excluded participant marked as such, are openly available in the Radboud Data Repository collection Interactive Semantic Segmentation for Phosphene Vision Neuroprosthetics at \url{https://doi.org/10.34973/m6kb-ec11}. The analysis and preprocessing code, the task and simulator code used in the experiment and the precomputed segmentation masks are available from the corresponding author upon request. The stimulus images are available from the Object Clutter Indoor Dataset under its own terms.}

\section*{Ethical statement}

The study was approved by the Ethics Committee Social Sciences (ECSW) of Radboud University (approval number ECSW-2025-036). All participants gave written informed consent prior to participation, in accordance with the Declaration of Helsinki.

\section*{Conflicts of interest}

The authors declare that they have no competing interests.

\suppdata{Supplementary material (PDF) in nine sections, the Canny parameter sweep (S1), chance levels for the combined accuracy measure (S2), model validation (S3), prior specifications (S4), convergence diagnostics (S5), prior sensitivity (S6), interval widths (S7), gaze contact at alternative tolerances (S8) and the signal detection analysis (S9).}

\bibliographystyle{unsrt}

\bibliography{refs}

\end{document}